%% file: capitanio.tex
\documentclass{PoS}
\newcommand{\bhc}{4U~1630-472\  }
\newcommand{\bhcc}{4U~1630-472.\  }

\title{The non standard evolution of the Compton corona in the three 2006-2010 subsequent outbursts of the Black Hole Candidate 4U 1630-47}

\ShortTitle{The non standard evolution of the Compton corona of the BHC 4U 1630-472}

\author{ \speaker{Fiamma Capitanio}%
         \thanks{The
RXTE and the XRT data were obtained through the HEASARC online
service. The Swift/BAT light curve is part of the SWIFT/BAT transient
monitor results provided by the Swift/BAT team. The INTEGRAL data were obtained through the ISDC online service. }\\
        INAF/IAPS via Fosso del Cavaliere, 100, 00133 Rome, Italy\\
        E-mail: \email{fiamma.capitanio@iasf-roma.inaf.it}}
\author{Riccardo Campana\\
        INAF-IASF, via Piero Gobetti, 101 - 40129 Bologna, Italy\\
        }

\author{Giovanni De Cesare\\
       INAF-IASF, via Piero Gobetti, 101 - 40129 Bologna, Italy\\
}
\author{Carlo Ferrigno\\
        ISDC, Data Centre for Astrophysics, Chemin d'Ecogia 16 CH-1290 Versoix, Switzerland\\
}
\abstract{We report on the analysis of the data collected by Swift, INTEGRAL and RXTE  of the Black Hole Candidate (BHC) \bhc during 3 consecutive outbursts occurred in 2006, 2008 and 2010, respectively. We show that, although a similar spectral and temporal behaviour in the energy range between 2-10 keV, these 3 outbursts present pronounced differences above 20 keV.  In fact, the 2010 outburst extends at high energies without any detectable cut-off until 150-200 keV, while the other two previous outbursts, occurred on 2006 and 2008, are not detected at all above 20 keV.  Moreover, the 2008 outburst does not show any detectable hard state in its final phases and even during the 2010 outburst, the  final hard state shows some peculiarities rarely observed in other BHC. 
We also investigate on the peculiar huge variation of \bhc hydrogen column density (N$_{H}$) reported in the literature using the Swift/XRT data. In fact this instrument is one of the most suitable for this purpose thanks to its lower energy coverage.
}

\FullConference{Swift: 10 Years of Discovery\\
                 2-5 December 2014\\
                 La Sapienza University, Rome, Italy}

\begin{document}

\section{Short history of the source}
\bhc underwent an outburst at least 20 times since 1969 and has been observed by several X-ray missions collecting a huge amount of data. \bhc is an highly absorbed source, $N_{H} = (5-12)\times10^{22} cm^{-2}$ \cite{Tomsick98} and the discovery of two dips in the light curve implies that probably the inclination angle is greater than 60$^{\circ}$~\cite{Tomsick98,Tomsick05}. The distance and the mass of the compact object have not a direct measurement because no optical counterpart has been identified until now. 
However, the infrared studies confirmed that the source lies in a crowded region in the direction of a giant molecular cloud located at 11 kpc, but in the near side of this cloud (thus at a distance $\leq$11 kpc)~\cite{Augusteijn}.
Moreover the infrared properties are consistent with a relatively long orbital period system ($\sim$few days) containing an early type companion~\cite{Augusteijn}. 

\section{Data reduction}

During the 2006-2010 outbursts of \bhc more than 800 ks of observations have been collected by both the RXTE Proportional Counter Array (PCA) and the INTEGRAL $\gamma$-ray telescope IBIS.
 The RXTE/PCA and INTEGRAL/IBIS data analysis was performed following the data reduction software as described in~\cite{Capitanio15} and reference therein. As previously reported by ~\cite{Tomsick14} and ~\cite{Capitanio15}, the PCA data of \bhc could be contaminated by the nearby  X-ray source IGR J16320-4751, that is a well studied high mass X-ray binary (HMXRB) lying at  0.25$^\circ$ away from \bhcc For this reason the PCA observations collected during the final phases of the 3 outbursts with less than 50 counts/s cannot be considered totally reliable~\cite{Capitanio15}. As reported by~\cite{Tomsick05}, there is another HMXRB that could contaminated the spectra, IGR J16318-4848, but it was not active during the 3 outbursts of \bhc~\cite{Capitanio15}.
   Unfortunately, the XRT observations were  just concentrated during the final phases of the 2008 and 2010 outbursts, when also the INTEGRAL observation campaign was near to the end of the observations. For these reasons we were able to extract only one joint  XRT-PCA spectrum during 2008 outburst and a quasi simultaneous spectrum of XRT and IBIS during the 2010 one (the IBIS spectrum has a longer integration time respect to the XRT one).
In order to measure the source N$_{H}$, we analysed also the XRT data collected during 2012 outburst.
The Swift/XRT data analysis was performed following the standard procedures~\cite{Burrows}. For the spectral analysis we considered only the XRT data observed in window timing mode (WT) in order to have a good spectral resolution.

\section{Results and Discussion}

\begin{figure*}
\centering
\includegraphics[scale=0.47,angle=90]{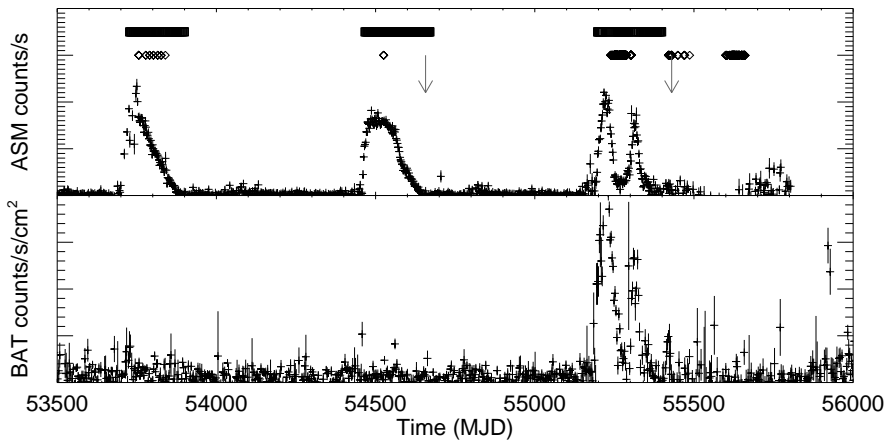}

\caption{Top panel: \bhc ASM light curve (bin size 2 d). The rectangles and the diamonds represent the PCA and INTEGRAL observation dates, respectively.  The two arrows represent the observation dates of the XRT spectra of Figure~2. Bottom panel: 15--50 keV Swift/BAT light curve (bin size 3 d). BAT 1$\sigma$  detection sensitivity is 5.3 mCrab for a full-day observation; 1mCrab is 0.00022 cnts$\times$cm$^{-2}$s $^{-1}$ [2].}
\label{lcurve}
\end{figure*}

\begin{figure*}
\centering
\includegraphics[scale=0.5,angle=0]{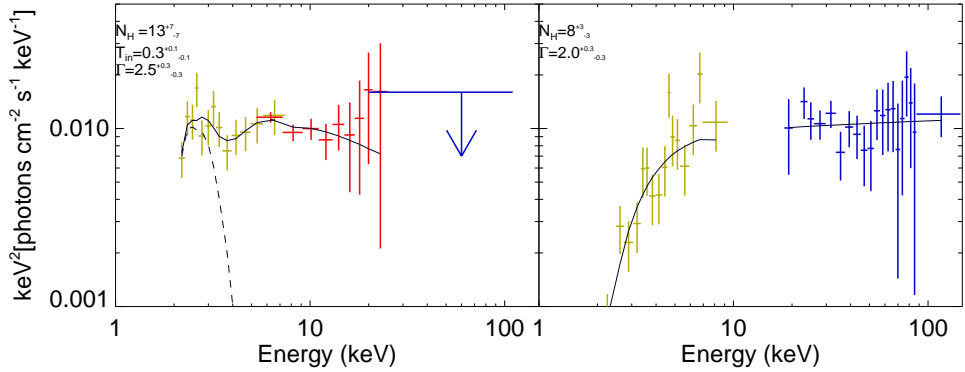}
\caption{Left panel: simultaneous XRT (green points)-PCA(red points) spectrum during the final part of the \bhc
 2008 outburst (MJD=54657). In blue the 2008 IBIS upper limit. The model used is an absorbed multi-color disk black body~\cite{Sakura}  plus a power law. A normalization constant was added in order to consider the different instrument calibrations. 
  Right panel: quasi simultaneous XRT (green points)-IBIS (red points) spectrum during the final phase of the 2010 outburst (XRT MJD$\sim$55430, IBIS integration interval: 55420--55486 MJD. The model consists in a simple absorbed power law plus a constant to take into account the the different instruments calibrations and integration time. The black lines of the two figures represent the model components.}

\label{f2008-2010}
\end{figure*}

As Figure~\ref{lcurve} shows, the behaviour in the 2--20 keV energy range is basically the same for the first two outbursts, while the third outburst evolves at harder energies and presents two bright peaks in the RXTE/All-Sky Monitor (ASM) and BAT light curves, respectively. Thus, the first two outbursts do not show any emission above 20 keV as the Swift/BAT light curve shows (Figure~\ref{lcurve} bottom panel). On the contrary the BAT light curve of the 2010 outburst is at a level flux about 14 times greater than the 2006 one. The detailed spectral evolution reported by ~\cite{Capitanio15} shows that during the 2006 and 2008 outbursts a power law component, even if very faint, is always necessary in order to obtain a good fit even in the softest spectra. Moreover the lack of BAT and IBIS detection implies that there should be a cutoff in the spectra at about 30 keV. 
The left panel of Figure~\ref{f2008-2010}  shows a XRT-PCA spectrum of the final phase of the 2008 outburst plus the IBIS upper limit, while right panel of Figure ~\ref{f2008-2010} shows a XTE-IBIS spectrum of the 2010 outburst (the PCA data are too contaminated by the nearby source to be used). Both spectra have been collected near the end of the two outbursts (see the arrows in Figure~\ref{lcurve}).
The 2008 spectrum even if very faint shows a shape typical of a soft state. Whereas  the 2010 spectrum seems to have reached the hard state, as previously reported by~\cite{Tomsick14}, because of the lack of the black body component. However, the power law photon index is instead typical of a soft state (see Figure~\ref{f2008-2010}).
As reported by~\cite{Capitanio15}, this source  in 2006 and 2008 undergone an outburst without showing an initial hard state. Our results also show that in 2008 probably did not show an hard state even in the final part of the outburst. As reported by~\cite{Tomsick14},  the final transition at the end of the 2010 outburst instead was delayed. However, the joint XRT-IBIS spectrum reveals that the Compton corona has characteristics more similar to a soft state than to a hard state.  However,  the interpretation of these data is more difficult because of the large IBIS integration time and the faintness of the spectrum. 

\section{Swift/XRT N$_{H}$ measurement of \bhc: preliminary results}
Several values of N$_{H}$, not consistent to each other, have been reported in literature ($\Delta$N$_{H}$=(4-12) $\times$10$^{22}$ cm$^{-2 }$, see e.g.~\cite{Tomsick98}).
We investigated if this variation (reaching 30\%) is real or is due to the different model used in the fitting procedure. For this purpose we analysed several observations of the source collected during the various outbursts in WT mode. { We selected those in which the source shows a prominent disk black body emission in order to avoid adding the power law to the spectral model that could interfere with the absorption modelization at low energies}.  
Table~\ref{tab} shows our results:  a slight variation of the N$_{H}$ value is present in the data ($\sim$9\%), but actually lower than the one extrapolated from literature. Even the N$_{H}$ values (affeced by huge errors) extracted from the two spectra in Figure~\ref{f2008-2010} are consistent with the values reported in Table~\ref{tab}.
However,  to obtain a complete sample of results,  a more coordinated observation campaign of Swift and INTEGRAL should be performed in order to cover all the outburst phases. In fact, while XRT could well modelize the absorption and the accretion disk, {IBIS could instead modelize the high energy part of the spectra using comptonization models instead of power law 
in order to minimize the interference at the low energy range.}

\begin{table*}
\centering

\caption{Fit results of 4 XRT observations of \bhc with a prominent emission of the black body component. the model used for the fit is an absorbed multicolor disk black body~\cite{Sakura}.
 }
\label{tab}

\resizebox*{0.6\textwidth}{!}{\begin{tabular}{lcccccccccc}

\hline
\hline

\input{tab1.tex}

\hline
\hline
\end{tabular}
}

\end{table*}

\end{document}

%% file: tab1.tex
XRT ID & N$_{H}$ & T$_{in}$ & N$_{disk}$ & Flux (2--10 keV) \\
-- &  cm$^{-2}$& KeV & -- & ergs$\times$cm$^{-2}\times$s$^{-1}$\\
00400107000 & 9.4$\pm$0.2 & 1.5$\pm$0.1 & 192$^{+34}_{-29}$ & 1.2$\times$10$^{8}$\\
00031224010 & 8.8$\pm$0.1 & 1.87$\pm$0.03& 121$\pm$8 & 2.0$\times$10$^{8}$\\
00521085000 & 9.1$\pm$0.1 & 1.83$\pm$0.02& 117$\pm$7 & 1.8$\times$10$^{8}$\\
00521085001& 9.1$\pm$0.1 & 1.83$\pm$0.02& 148$\pm$9 & 2.3$\times$10$^{8}$\\